\documentclass[doublecol]{epl2} 

\usepackage{amsfonts,amsmath}

\usepackage{epsfig}
\usepackage{color}
\usepackage{bm}

\usepackage{braket}

\usepackage{graphicx}

\newcommand{\eq}[1]{\begin{equation} #1 \end{equation}}
\newcommand{\eqa}[2]{\begin{equation} #1 \label{#2} \end{equation}}
\newcommand{\balign}[1]{\begin{align} #1 \end{align}}

\newcommand{\figin}[4]
{\begin{figure}[tb]
\centering
\includegraphics[width= #1]{#2.pdf}
\caption{#3}
\label{f:#4}
\end{figure}}

\newcommand{\todayd}{\the\year/\the\month/\the\day}

\newcommand{\bib}{\bibitem}

\newcommand{\up}{\uparrow}

\newcommand{\lb}{\label}
\newcommand{\nt}{\notag}
\newcommand{\Tr}{\mathrm{Tr}}

\newcommand{\bel}{\begin{easylist}}
\newcommand{\eel}{\end{easylist}}

\newcommand{\eref}[1]{Eq.~\eqref{#1}}

\newcommand{\fref}[1]{Fig.~\ref{f:#1}}

\def \({\left(}
\def \){\right)}
\def \[{\left[}
\def \]{\right]}
\newcommand{\la}{\left\langle}
\newcommand{\ra}{\right\rangle}

\newcommand{\sumtwo}[2]%
{\mathop{\sum_{#1}}_{#2}}
\newcommand{\sumthree}[3]%
{\mathop{\mathop{\sum_{#1}}_{#2}}_{#3}}
\newcommand{\sumfour}[4]%
{\mathop{\mathop{\mathop{\sum_{#1}}_{#2}}_{#3}}_{#4}} 
\newcommand{\prodtwo}[2]%
{\mathop{\prod_{#1}}_{#2}}
\newcommand{\mintwo}[2]%
{\mathop{\min_{#1}}_{#2}}
\newcommand{\maxtwo}[2]%
{\mathop{\max_{#1}}_{#2}}
\newcommand{\maxthree}[3]%
{\mathop{\mathop{\max_{#1}}_{#2}}_{#3}}
\newcommand{\limtwo}[2]%
{\mathop{\lim_{#1}}_{#2}}
\newcommand{\suptwo}[2]%
{\mathop{\sup_{#1}}_{#2}}
\newcommand{\supthree}[3]%
{\mathop{\mathop{\sup_{#1}}_{#2}}_{#3}}
\newcommand{\supfour}[4]%
{\mathop{\mathop{\mathop{\sup_{#1}}_{#2}}_{#3}}_{#4}} 
\newcommand{\inftwo}[2]%
{\mathop{\inf_{#1}}_{#2}}
\newcommand{\infthree}[3]%
{\mathop{\mathop{\inf_{#1}}_{#2}}_{#3}}
\newcommand{\inffour}[4]%
{\mathop{\mathop{\mathop{\inf_{#1}}_{#2}}_{#3}}_{#4}} 

\newcommand\calN{{\cal N}}

\newcommand{\bsA}{\boldsymbol{A}}
\newcommand{\bsB}{\boldsymbol{B}}
\newcommand{\bsC}{\boldsymbol{C}}
\newcommand{\bsD}{\boldsymbol{D}}

\newcommand{\para}[1]{{\em #1}\/.---}

\def\x{\overline{X}}
\def\y{\overline{Y}}
\def\z{\overline{Z}}
\def\a{\overline{A}}

\def\bz{\overset{Z}{|}}
\def\hz{\underset{Z}{\up}}

\newcommand{\qh}{$[Q, H]$ }

\newcommand{\lplus}{\overset{\to}{+}}
\newcommand{\rplus}{\overset{\leftarrow}{+}}

\def\i{{\rm i}}

\newcommand{\ba}[2]{
\begin{array}{#1}
#2
\end{array}
}

\newcommand{\rev}{}

\def\rnum#1{\resizebox{0.5em}{\height}{\expandafter{\romannumeral #1}}}
\def\Rnum#1{\resizebox{0.5em}{\height}{\uppercase\expandafter{\romannumeral #1}}}

\renewcommand{\labelenumi}{(\roman{enumi})}

\newcommand{\titlename}{Proof of the absence of local conserved quantities in the XYZ chain with a magnetic field}

\title{\titlename}

\author{Naoto Shiraishi}
\institute{Department of Physics, Gakushuin University, 1-5-1 Mejiro, Toshima-ku, Tokyo, 171-8588, Japan}%


\date{\today}

\abstract{
We rigorously prove that the spin-1/2 XYZ chain with a magnetic field has no local conserved quantity.
Any nontrivial conserved quantity of this model is shown to be a sum of operators supported by contiguous sites with at least half of the entire system.
{We establish that the absence of local conserved quantity in concrete models is provable in a rigorous form.}}

\begin{document}


\maketitle

\section{Introduction}
Integrable many-body systems have played momentous roles in various research fields, including statistical mechanics~\cite{JM, Bax, Tak}, strongly-correlated systems~\cite{LW, And, Wie, KM}, quantum computation~\cite{Val, Kni, TD}, and high-energy physics~\cite{MNS, MZ, BPR}.
Although there is no established definition of quantum integrability unlike classical integrability~\cite{Wei, YS, CM, GE16}, for locally-interacting many-body spin systems the integrability is roughly equivalent to the existence of infinitely-many local conserved quantities, which guarantees exact solutions of these systems~\cite{Bax, Tak, DG, FF, GM94, GM95, KBI}.
The presence or absence of local conserved quantities is relevant to various aspects of many-body systems.
For example, systems with local conserved quantities do not thermalize to the standard equilibrium state~\cite{RDYO, Caz, Lan, EF}, while systems without local conserved quantities appear to thermalize~\cite{Rig08, Sor, GE16, SM, MS, Shi17}.
Another example is the energy level statistics, which obeys the Poisson distribution in integrable systems and obeys the Wigner-Dyson distribution in chaotic systems~\cite{Haa}.
The Bethe ansatz and the quantum inverse scattering method are useful to find out energy eigenstates and many local conserved quantities in integrable systems~\cite{Bax, Fad, KBI}.
Now various integrable models, including the Heisenberg spin-1/2 chain, the XYZ chain, the IRF model, and many other more complicated models, have been discovered, and their classification and characterization have been investigated~\cite{ABF, AKW, OY, YS}.

Although vast literature is devoted to integrability, very few studies have addressed non-integrability of specific models in spite of its necessity.
Here, we used the word {\it non-integrable} in the sense that the model has no local conserved quantity.
\rev{Non-integrability is of course relevant to the research field of chaos and thermalization, but it is also relevant to much more broad research fields.
Non-integrability is strongly related to the presence of thermalization, and thus if one wants to avoid anomalous non-thermalizing phenomena, they should ensure the non-integrability of the system in consideration (see also \fref{table}).
For instance, a kind of ergodic property is employed in the derivation of the Kubo formula~\cite{Kub, SF, SK}, and some integrable systems obey a modified version of the Kubo formula~\cite{Suz}.
In many other situations from local equilibration in hydrodynamic description~\cite{HHNH} to scrambling in black holes~\cite{KMS, Las, SS14, IS18}, non-integrability of systems in consideration is explicitly or implicitly assumed.
Furthermore, transport properties are also affected by the integrability of systems in consideration~\cite{Zot}.
These facts confirm that proving non-integrability of concrete models in a rigorous form is relevant to broad research fields.}
However, as stated above, very few theoretical works have tackled to show the non-integrability of a certain model.
A notable exception is Ref.~\cite{GM95-2}, which tries to argue non-integrability of certain specific models by putting some hypotheses.
However, this attempt only draws a heuristic road map, and a rigorous proof of non-integrability has been completely elusive.
Some researchers even have a pessimistic view that non-integrability of a specific model is out of the scope of analytical approach, and it can only be presumed with numerical simulations.

{To break this pessimistic view,} in this Letter, we provide the first rigorous proof of the absence of local conserved quantities in a specific model, the one-dimensional $S=1/2$ XYZ spin systems with a magnetic field.
We show that any nontrivial conserved quantity is a sum of operators supported by contiguous sites with at least half of the system, which means non-locality of this conserved quantity.
Our strategy is straightforward.
We first write down a formal expansion of a candidate of local conserved quantities consisting of products of at most $k$ neighboring spin operators, and demonstrate that if it conserves, then all of the coefficients in this expansion must be zero.
The case with $k=3$, which is treated in detail, already contains the essence for general $k$.
Our approach to proving non-integrability can extend to other spin systems.
{This result opens a novel research direction where non-integrability is investigated in a mathematically rigorous form.}

\rev{

\figin{8.5cm}{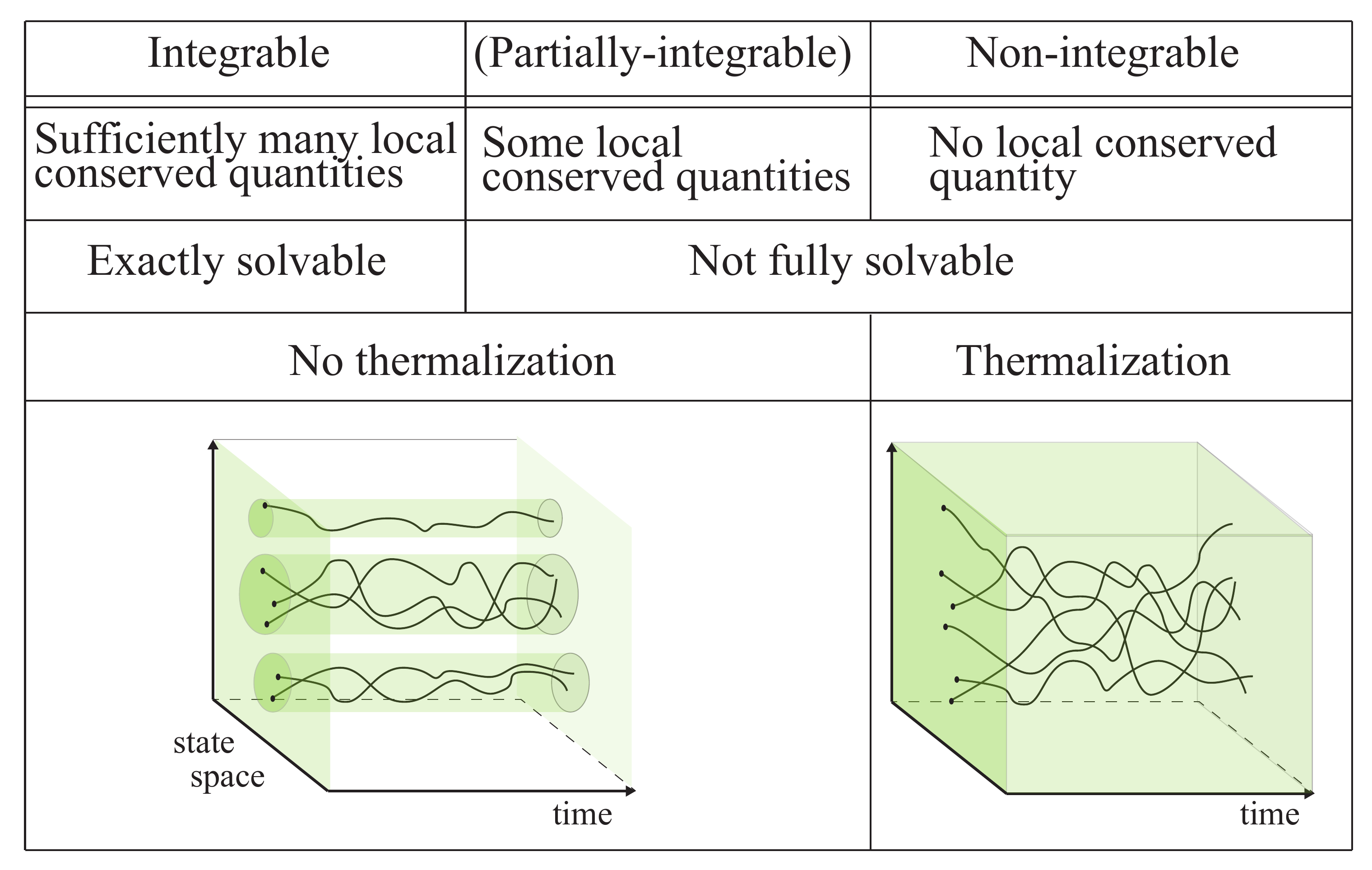}{
We employ a working definitions of quantum integrability and non-integrability: the presence and absence of local conserved quantities.
Integrable systems have sufficiently many local conserved quantities, and their eigenenergies and eigenstates are exactly solvable.
Non-integrable systems have no local conserved quantities.
Systems with local conserved quantities do not thermalize, and systems with no local conserved quantity are considered to thermalize.
Two schematics draw the role of local conserved quantities in the time-evolution of states.
If a system has local conserved quantities, the system can evolve only in subspace with fixed local conserved quantities (the tubes in the left figure).
Although almost all physical systems in nature are considered to be non-integrable, maybe surprisingly no concrete model has been proven to be non-integrable.
}{table}
}

\section{Setup and main claim}

We consider the standard $S=1/2$ XYZ spin chain with $L$ sites under a magnetic field with the periodic boundary condition:
\eqa{
H=-\sum_{i=1}^L [ J_X X_i X_{i+1}+J_Y Y_iY_{i+1}+J_ZZ_iZ_{i+1}] -\sum_{i=1}^L h Z_i, \nt
}{XYZ+h}
where $X$, $Y$, $Z$ represents the Pauli matrices $\sigma^x$, $\sigma^y$, $\sigma^z$, and we set all the coupling constants $J_X$, $J_Y$, $J_Z$ nonzero.
Since the Hamiltonian is translationally invariant, we can write any conserved quantities in a translationally invariant form  (see \cite{supple}).
We introduce a symbol $\bsA^l_i:=A^1_iA^2_{i+1}\cdots A^l_{i+l-1}$, which is a sequence of $l$ operators starting from the site $i$ to the site $i+l-1$.
The first and the last operators $A^1$ and $A^l$ take one of the Pauli operators ($X$, $Y$, or $Z$), and other operators $A^2,\cdots ,A^{l-1}$ take one of the Pauli operators or the identity operator $I$.
Using this symbol, we write a candidate of local conserved quantities as
\eqa{
Q=\sum_{l=1}^{k} \sum_{\bsA^l} \sum_{i=1}^L q_{\bsA^l}\bsA^l_i
}{Qform}
with coefficients $q_{\bsA^l}$.
The sum of $\bsA^l$ runs over all possible sequences from $XX\cdots XX$ to $ZI\cdots IZ$.
Since the Pauli matrices and the identity span the space of $2\times2$ Hermitian matrices, we find that the above form covers all possible translationally invariant quantities.
We call $\sum_i \bsA^l_i$ an {\it $l$-support operator}.
For example, there are $3\times4\times3=36$ possible 3-support operators.
If one of $q_{\bsA ^k}$ is nonzero and all $q_{\bsA^l}$ with $l\geq k+1$ are zero, we call $Q$ as a {\it $k$-support conserved quantity}.
{For example, the Hamiltonian itself is a 2-support conserved quantity, and the projection operator onto an energy eigenstate is usually an $L$-support conserved quantity.}
The claim of this Letter is that the spin chain \eqref{XYZ+h} has no $k$-support conserved quantity with $3\leq k\leq L/2$ as long as $J_X\neq J_Y$ and $h\neq 0$.
In other words, all nontrivial conserved quantities are highly nonlocal.

\section{Strategy}
Take a $k$-support operator $Q$ in the form \eqref{Qform}.
We consider the commutator of $Q$ and $H$:
\eqa{
[Q,H]=\sum_{l=1}^{k+1}\sum_{\bsB^l} \sum_{i=1}^L r_{\bsB^l} \bsB^l_i,
}{QH}
which is an at most $k+1$-support operator because $Q$ is a $k$-support operator and $H$ is a 2-support operator.
The conservation of $Q$ reads $r_{\bsB^l}=0$ for any $\bsB^l$, which results in a system of linear relations on $q_{\bsA}$ by comparing both sides of \eref{QH}.
We shall show that these linear relations do not have nontrivial solutions except that with all $q_{\bsA ^k}$ zero.
To prove this, we narrow the range of possible coefficients $q_{\bsA ^k}$ which may take nonzero values step by step, and finally ensure that no coefficient can take nonzero value with keeping all relations consistent.

Our proof consists of two steps.
\begin{enumerate}
\renewcommand{\labelenumi}{\alph{enumi}.}
\item In the first step, we exploit the condition $r_{\bsB^{k+1}}=0$ for all $\bsB^{k+1}$, and show that (i) the coefficients of $\bsA^k$ except those in a specific form are zero, (ii) all the remaining coefficients of $\bsA^k$ are in the linear relation.
\item In the second step, we exploit the condition $r_{\bsB^k}=0$ for all $\bsB^k$, and show that one of the remaining coefficient of $\bsA^k$ is zero.
\end{enumerate}
Owing to the linear relation shown in a-(ii), this completes the proof of absence of $k$-support conserved quantities.

Since the case with $k=3$ captures the essential idea of the proof for general $k$, in the following we describe the proof for the case with $k=3$ in detail, and then outline the proof for general $k$.

\section{Case of $k=3$: Step 1}
We consider a 3-support conserved quantity $Q$ in this and next section.
In the first step, we focus on 4-support operators in $[Q,H]$.
In this letter, if a commutator with an operator $\bsA$ and $\bsC$ results in another operator $\bsD$, we say that the operator $\bsD$ is {\it generated} by the commutator with $\bsA$ and $\bsC$.
In case of $k=3$, a 4-support operator in $[Q,H]$ is generated by commutators between a 3-support operator in $Q$ and a 2-support operator in $H$.
For example, the 4-support operator $\sum_i X_iY_{i+1}Y_{i+2}X_{i+3}$ in \qh is generated by the following two commutators:
\balign{
-\i [X_iY_{i+1}Z_{i+2}, X_{i+2}X_{i+3}]&=2X_iY_{i+1}Y_{i+2}X_{i+3}, \\
-\i [Z_{i+1}Y_{i+2}X_{i+3}, X_iX_{i+1}]&=2X_iY_{i+1}Y_{i+2}X_{i+3}.
}
In case without confusion, we drop the summation of $i$ and subscripts for visibility.
We express these two relations as
\eq{
\begin{array}{rcccc}
&X&Y&Z& \\
&&&X&X \\ \hline
2&X&Y&Y&X
\end{array} 
\ \ \ \ \ 
\begin{array}{rcccc}
&&Z&Y&X \\
&X&X&& \\ \hline
2&X&Y&Y&X,
\end{array} \nt 
}
where the horizontal bar represents a commutator (including the imaginary number $\i$), and the horizontal positions correspond to the positions of spin operators.
With noting that the operator $XYYX$ in $[Q,H]$ is generated only by the above two commutators, the condition $r_{XYYX}=0$ implies
\eq{
q_{XYZ}+q_{ZYX}=0.
}

Notably, some 4-support operators are generated only by a single commutator.
An example is $XXXY$, which is generated only by
\eqa{
\begin{array}{rcccc}
&X&X&Z& \\
&&&Y&Y \\ \hline
-2&X&X&X&Y.
\end{array}
}{XXXY}
This fact directly means
\eq{
q_{XXZ}=0.
}
In addition, by considering commutators
\eq{
\begin{array}{rcccc}
&Y&Z&Y& \\
&&&Z&Z \\ \hline
2&Y&Z&X&Z
\end{array} 
\ \ \ \ \ 
\begin{array}{rcccc}
&&X&X&Z \\
&Y&Y&& \\ \hline
2&Y&Z&X&Z,
\end{array} \nt
}
we have
\eq{
J_Zq_{YZY}=-J_Yq_{XXZ}=0.
}
In a similar manner, we arrive at the fact that $q_{ABC}=0$ if two of $A$, $B$, $C$ are the same.
Moreover, we have $q_{AIB}=0$ for any $A$, $B$, which follows from 
\eq{
\begin{array}{rcccc}
&X&I&Z& \\
&&&Y&Y \\ \hline
-2&X&I&X&Y
\end{array} \nt
}
and similar relations.

In summary, the analysis of 4-support operators in \qh yields
\balign{
&J_Xq_{YXZ}=J_Yq_{ZYX}=J_Zq_{XZY} \nt \\
=&-J_Xq_{ZXY}=-J_Yq_{XYZ}=-J_Zq_{YZX}, \lb{4-support-sum}
}
and coefficients of other 3-support operators in $Q$ turn out to be zero.

\section{Case of $k=3$: Step 2}
We next focus on 3-support operators in $[Q,H]$.
First, $YZY$ is generated by the following four commutators;
\balign{
\begin{array}{rccc}
&Y&Z&X \\
&&&Z \\ \hline
-2&Y&Z&Y
\end{array} 
&\ \ \ \ \
\begin{array}{rccc}
&X&Z&Y \\
&Z&& \\ \hline
-2&Y&Z&Y
\end{array} 
\nt \\
&\nt \\
\begin{array}{rccc}
&Y&X& \\
&&Y&Y \\ \hline
2&Y&Z&Y
\end{array} 
&\ \ \ \ \
\begin{array}{rccc}
&&X&Y \\
&Y&Y& \\ \hline
2&Y&Z&Y,
\end{array} \nt
}
which reads
\eqa{
h(q_{YZX}+q_{XZY})-J_Y(q_{YX}+q_{XY})=0.
}{3-YZY}

Next, considering $YYZ$
\eq{
\begin{array}{rccc}
&X&Y&Z \\
&Z&& \\ \hline
-2&Y&Y&Z
\end{array} 
\ \ \ \ 
\begin{array}{rccc}
&Y&X&Z \\
&&Z& \\ \hline
-2&Y&Y&Z
\end{array}
\ \ \ \ 
\begin{array}{rccc}
&Y&X& \\
&&Z&Z \\ \hline
-2&Y&Y&Z
\end{array}  \nt
}
and $XXZ$
\eq{
\begin{array}{rccc}
&X&Y&Z \\
&&Z& \\ \hline
2&X&X&Z
\end{array} 
\ \ \ \ 
\begin{array}{rccc}
&Y&X&Z \\
&Z&& \\ \hline
2&X&X&Z
\end{array}
\ \ \ \ 
\begin{array}{rccc}
&X&Y& \\
&&Z&Z \\ \hline
2&X&X&Z,
\end{array}  \nt
}
both of which are generated by three commutators, we further have
\balign{
h(q_{XYZ}+q_{YXZ})+J_Zq_{YX}=&0, \lb{3-YYZ} \\
h(q_{XYZ}+q_{YXZ})+J_Zq_{XY}=&0. \lb{3-XXZ}
}
Combining Eqs.~\eqref{3-YZY}, \eqref{3-YYZ}, and \eqref{3-XXZ} to erase $q_{YX}$ and $q_{XY}$, along with relations $q_{YZX}=-q_{XZY}$ and $J_Xq_{YXZ}=-J_Yq_{XYZ}$ shown in \eref{4-support-sum}, we arrive at
\eq{
h\( 1-\frac{J_Y}{J_X}\) q_{XYZ}=0. 
}
Hence, $q_{XYZ}=0$ holds as long as $h\neq 0$ and $J_X\neq J_Y$. 
Due to \eref{4-support-sum}, $q_{XYZ}=0$ suffices to prove the absence of 3-support conserved quantities in the XYZ chain with a magnetic field.

\section{General case: Step 1}
We proceed to analyses on $k$-support conserved quantities $Q$ with general $3\leq k\leq L/2$~\cite{note-k}.
We shall show that such a $Q$ does not exist.

In the first step, we focus on $k+1$-support operators in \qh.
To explain our findings, we introduce a useful expression of operators such as
\balign{
\x\y\x _i&=c (X_iX_{i+1})(Y_{i+1}Y_{i+2})(X_{i+2}X_{i+3}) \nt \\
&=X_i Z_{i+1} Z_{i+2}X_{i+3},
}
where $c$ takes one of $\{ \pm1, \pm \i \}$ to make its coefficient 1.
The symbol $\a$, which we call {\it doubling product}, represents the exchange interaction of $A$, and the neighboring symbol has its support with single-site shift.
The coefficient $c$ leads to the rule of the coefficientless product of the Pauli operators: $XY=YX=Z$, $YZ=ZY=X$, $XZ=ZX=Y$.
We require that the same symbols cannot be neighboring (e.g., $\x\x\z$ is not allowed).
If an operator can be expressed in the above form, we call this operator as {\it doubling-product operator}.

We now claim that the condition $r_{\bsB^{k+1}}=0$ for all $\bsB^{k+1}$ leads to the following two facts: 
\begin{enumerate}
\item All $k$-support operators in $Q$ which are not doubling-product have zero coefficients, 
\item Any two coefficients of doubling-product $k$-support operators in $Q$ have a linear relation as \eref{4-support-sum}.
\end{enumerate}
In case of $k=3$, there are six 3-support operators which are doubling-product; $\x\y$, $\x\z$, $\y\x$, $\y\z$, $\z\x$, $\z\y$, and they are the six operators appearing in \eref{4-support-sum}.
The aforementioned claims ares their generalization.

The latter fact (ii) is ensured as follows:
Take the case of $YZZZYZ$ ($k=5$) as an example.
Two commutators
\balign{
&\begin{array}{rcccccc}
&Y&Z&Z&Z&X& \\
&&&&&Z&Z \\ \hline
-2&Y&Z&Z&Z&Y&Z
\end{array}
\nt \\
& \nt \\
&\begin{array}{rcccccc}
&&X&Z&Z&Y&Z \\
&Y&Y&&&& \\ \hline
2&Y&Z&Z&Z&Y&Z
\end{array}  \nt
}
imply that $q_{YZZZX}$ and $q_{XZZYZ}$ have a linear relation.
The reason why $YZZZX=\y\x\y\x$ and $XZZYZ=\x\y\x\z$ have a relation is that $\y\x\y\x$ becomes $\x\y\x\z$ by adding $\z$ from right and removing the leftmost $\y$.
In general, if an operator is obtained from another operator by removing the leftmost (rightmost) doubling operator and adding a doubling operator from right (left), then the coefficients of these two operators have a linear relation.
This observation directly leads to the latter fact (ii).

The former fact (i) is ensured in a similar line to above.
If an operator is not doubling-product, this operator has at least one {\it inconsistency} in the doubling-product representation.
Consider $XYZZX=\x\z \cdot (ZX)$ as an example.
By removing the leftmost $\x$ and adding $\y$ from right as in the above paragraph, the obtained operator $ZZZZY$ obviously has zero coefficient for the same reason as \eref{XXXY}.
In a similar manner to above, all non-doubling-product operator can be shown to be zero.

\section{General case: Step 2}
From the facts (i) and (ii), it suffices to prove one of the coefficient of a doubling-product $k$-support operators in $Q$ zero, which is accomplished by considering $k$-support operators in $[Q,H]$.
Similar to the case with $k=3$, a $k$-support operator in \qh is generally generated by four commutators; two commutators between a $k$-support operator in $Q$ and a magnetic field in $H$, and two commutators between a $k-1$-support operator in $Q$ and the exchange interaction in $H$.
For example, $ZXZXZ$ is generated by the following four commutators:
\balign{
\ba{rccccc}{
&Z&X&Z&Y&Z \\
&&&&Z& \\ \hline
2&Z&X&Z&X&Z,
}
&\ \ 
\ba{rccccc}{
&Z&Y&Z&X&Z \\
&&Z&&& \\ \hline
2&Z&X&Z&X&Z,
}
\lb{ZXZXZ-1} \\ 
& \nt \\
\ba{rccccc}{
&Z&X&Z&Y& \\
&&&&Z&Z \\ \hline
2&Z&X&Z&X&Z,
}
&\ \ 
\ba{rccccc}{
&&Y&Z&X&Z \\
&Z&Z&&& \\ \hline
2&Z&X&Z&X&Z.
} \lb{ZXZXZ-2}
}
However, in some cases, a $k$-support operator in \qh is generated only by three commutators.
This happens when the two leftmost or rightmost operators of the $k$-support operator are the same, which we have already seen in case with $k=3$ (Eqs.~\eqref{3-YYZ} and \eqref{3-XXZ}).

To describe commutators as above in the doubling-product representation, we introduce some symbols.
First, we introduce a symbol ``$\hz$", which represents a commutator with $Z$ at this position.
For example, $\x\y\hz\z$ represents the commutator $[X_iZXZ, Z_{i+2}]$.
Here, the magnetic field $Z$ settles at the site $i+2$ because $\x\y\z=c(X_iX_{i+1})(Y_{i+1}Y_{i+2})(Z_{i+2}Z_{i+3})$ and the overlap of $\y$ and $\z$, which is referred by the upward arrow, is at the site $i+2$.
Using this symbol, the two commutators in \eqref{ZXZXZ-1} are expressed as
\eq{
\z\y\x\hz\z , \ \ \ \ \z\hz\x\y\z . \nt
}
Next, we introduce a symbol ``$\bz$", which represents multiplication of $Z$ at this position with setting coefficient 1.
Examples are $\z\x\bz=c(Z_iZ_{i+1})(X_{i+1}X_{i+2})Z_{i+2}=ZYY$ and $\x\bz\z\y\x=XXXZX$.
We note that the expression with $\bz$ is not unique (e.g., $ZXZY=\z\y\x\bz=\z\bz\x\y$).
We finally introduce two symbols ``$\rplus$" and ``$\lplus$", which mean commutators with the exchange interaction in $H$ at the rightmost and leftmost site, respectively.
Then, the two commutators in \eqref{ZXZXZ-2} are expressed as
\eq{
\z\y\x\bz \rplus \z , \ \ \ \ \  \z \lplus \bz\x\y\z . \nt
}

We now construct a sequence of sets of commutators for even $k\geq 6$, which is presented in the next page.
(In case of odd $k\leq 7$, we replace $\x\y \cdots \y \x$ in the sequence to $\y\x \cdots \y \x$.
The cases of $k=4$ and $k=5$ are treated separately in a similar manner~\cite{full}.).
\newcommand{\threeskip}{\ \ }
\begin{widetext}
\eq{
\ba{ccccccc}{
\hz \y\z \x\y \cdots \y \x\z\x\y &\threeskip& \x\hz\z \x\y \cdots \y \x\z\x\y &\threeskip&\bz \y\z \x\y \cdots \y \x\z\x \rplus \y &\threeskip& \\
\hz \x\y\z \x\y \cdots \y \x\z\x &\threeskip& \y\x\hz\z \x\y \cdots \y \x\z\x &\threeskip&\bz \x\y\z \x\y \cdots \y \x\z \rplus \x &\threeskip& \y \lplus \x\bz\z \x\y \cdots \y \x\z\x\\
\hz \y\x\y\z \x\y \cdots \y \x\z &\threeskip& \x\y\x\hz\z \x\y \cdots \y \x\z &\threeskip&\bz \y\x\y\z \x\y \cdots \y \x \rplus \z &\threeskip& \x \lplus \y\x\bz\z \x\y \cdots \y \x\z\\
\z\hz\y\x\y\z \x\y \cdots \y \x &\threeskip& \z\x\y\x\hz\z \x\y \cdots \y \x  &\threeskip&\z\bz\y\x\y\z \x\y \cdots \y \rplus \x &\threeskip& \z \lplus \x\y\x\bz\z \x\y \cdots \y \x \\
\vdots &\threeskip& \vdots &\threeskip& \vdots &\threeskip& \vdots\\
\vdots &\threeskip& \vdots &\threeskip& \vdots &\threeskip& \vdots\\
 \x\y \cdots \y \x\z\hz\y\x\y\z &\threeskip&  \x\y \cdots \y \x\z\x\y\x\hz \z&\threeskip&  \x\y \cdots \y \x\z\bz\y\x\y \rplus \z &\threeskip&  \x \lplus \y \cdots \y \x\z\x\y\x\bz \z \\
\z \x\y \cdots \y \x\z\hz\y\x\y &\threeskip& \z \x\y \cdots \y \x\z\x\y\x\hz &\threeskip& \z \x\y \cdots \y \x\z\bz\y\x \rplus \y &\threeskip& \z \lplus \x\y \cdots \y \x\z\x\y\x\bz \\
\y\z \x\y \cdots \y\x\z\hz\y\x &\threeskip& \y\z \x\y \cdots \y \x\z\x\y \hz &\threeskip& \y\z \x\y \cdots \y\x\z\bz\y \rplus \x &\threeskip& \y \lplus \z \x\y \cdots \y \x\z\x\y \bz \\
\x\y\z \x\y \cdots \y \x\z\hz\y &\threeskip& \x\y\z \x\y \cdots \y \x\z\x\hz &\threeskip& &\threeskip& \x \lplus \y\z \x\y \cdots \y \x\z\x\bz
} \nt
}
\end{widetext}

Here, $\x\y \cdots \y \x$ is the abbreviation of the alternation of $\x$ and $\y$.
Each horizontal row of the sequence consists of three or four commutators generating the same $k$-support operator in \qh, which yields a relation of coefficients.
For example, the first and second rows yield
\balign{
h(q_{\y\z \x\y \cdots \y \x\z\x\y}+q_{\x\z \x\y \cdots \y \x\z\x\y})& \nt \\
+J_Yq_{\bz \y\z \x\y \cdots \y \x\z\x}&=0 \\
& \nt \\
h(-q_{\x\y\z \x\y \cdots \y \x\z\x}+q_{\y\x\z \x\y \cdots \y \x\z\x}) &\nt \\
+J_Xq_{\bz \x\y\z \x\y \cdots \y \x\z}+J_Yq_{\x\bz\z \x\y \cdots \y \x\z\x}&=0.
}
Remarkably, any $k-1$-support operators in this sequence appears twice~\cite{same}.
By summing up corresponding relations with multiplying proper coefficients to cancel all the coefficients of $k-1$-support operators and using the linear relation for doubling-product operators obtained in the previous section, we arrive at
\eq{
h\( \frac{J_X}{J_Y}-1\) (k+2)q_{\y\z\x\y\cdots \y\x\z\x\y}=0.
}
This directly implies that all the coefficients of $k$-support operators are zero as long as $h\neq 0$ and $J_X\neq J_Y$.

\section{Discussion}

We have rigorously shown that there is no $k$-support conserved quantity in the XYZ chain with a magnetic field for $3\leq k\leq L/2$.
This result breaks the widespread pessimistic belief that non-integrability is out of the scope of analytical investigation and is only presumed with numerical supports.
Now a novel research direction to proving non-integrability in concrete many-body systems has opened.
The proposed techniques and ideas extend to other one-dimensional spin-1/2 systems such as those with next-nearest interaction, including the Majumdar-Ghosh model~\cite{MG} and the Shastry-Sutherland model~\cite{SS81}.
The extension to spin-1 systems such as the AKLT model~\cite{AKLT} looks not straightforward since the rule of the product of spin-1 operators is more complicated than the case of spin-1/2.
Extensions to frustration-free systems as explained above is especially important because recent studies have revealed that some frustration-free or frustration-free-like systems~\cite{CEM,Aro,Ber17,LM18, Shi19} have solvable nonthermal excited states in spite of its nonintegrability.
To highlight their significance, the proof of the non-integrability of these models is desirable.

Another challenging extension is to quasi-local conserved quantities{~\cite{def-quasilocal}}, whose importance has recently been discovered in some translationally invariant integrable systems, the Heisenberg chain and the XXZ chain~\cite{Pro, PI, PPSA, IMP, quasi-GGE, IMPZ}.
{Although the presence of quasi-local conserved quantities without local ones seems to be unlikely, our proof does not exclude the possibility of the presence of quasi-local ones in a rigorous sense.}
Since we have not yet fully understood the nature of quasi-local conserved quantities, the investigation for the absence of quasi-local conserved quantity is, unfortunately, out of reach at present.
After arriving at the full understanding, the extension to quasi-local conserved quantities will merit future investigation.

In closing this article, we comment on the relation to some previous results on integrable systems.
The local conserved quantities of the $S=1/2$ XYZ chain without a magnetic field has been obtained through the quantum inverse scattering method, and the explicit form of these local conserved quantities are calculated in Refs.~\cite{GM94, GM95}.
These local conserved quantities indeed satisfies the obtained relation \eqref{4-support-sum} and its generalizations for general $k$.
The quantum inverse scattering method provides a systematic way to construct local quantities consistent with the commutative condition with the Hamiltonian.
Algebraic structures including the Hopf algebra and the Virasoro algebra lie behind this construction~\cite{JM}.
Our result employs a complementary approach to this, where we demonstrate that no local quantity is consistent with the commutative condition with a certain class of Hamiltonian.
This fact, the absence of local conserved quantities, may also have algebraic characterizations, which will shed light on the general principle for presence and absence of local conserved quantities.

\para{Acknowledgement}
The author is grateful to Atsuo Kuniba for fruitful discussion.
The author thanks Sriram Shastry and Fabian Essler for informing some related works.
The author also thanks Hosho Katsura, Chihiro Matsui, Takashi Mori, Tomoyuki Obuchi, Akira Shimizu and Hal Tasaki for helpful comments.
This work was supported by Grant-in-Aid for JSPS Fellows JP17J00393.

\clearpage

\makeatletter
\long\def\@makecaption#1#2{{
\advance\leftskip1cm
\advance\rightskip1cm
\vskip\abovecaptionskip
\sbox\@tempboxa{#1: #2}%
\ifdim \wd\@tempboxa >\hsize
 #1: #2\par
\else
\global \@minipagefalse
\hb@xt@\hsize{\hfil\box\@tempboxa\hfil}%
\fi
\vskip\belowcaptionskip}}
\makeatother
\newcommand{\vo}{\upsilon}
\newcommand{\midskip}{\vspace{3pt}}

\setcounter{equation}{0}
\def\theequation{A.\arabic{equation}}

\pagestyle{empty}

\begin{widetext}

\begin{center}
{\bf \Large Supplementary Material for ``\titlename"}

\bigskip
Naoto Shiraishi
\end{center}

\bigskip

\begin{quotation}
In this Supplementary Material, we shall explain why we safely assume the form of the candidate for a conserved quantity as in the translationally invariant form.
\end{quotation}

\bigskip

\bigskip\noindent
{\bf \large Treatment of non-translationally-invariant candidates of $k$-support conserved quantities}
\midskip

As we have seen in the footnote \cite{supple} in the main text, a candidate for a conserved quantity $Q=\sum_{j=1}^s \sum_{x=1}^m C_{mj+x}^x$, which is not translationally invariant, reduces to the translationally invariant form $Q':=\sum_{x=0}^{m-1}T^{(x)}Q=\sum_{i=1}^L \( \sum_{x=1}^mC_i^x\)$.
If $Q'$ is a $k$-support operator, we can apply to $Q'$ the same analysis shown in the main text and conclude that such a $Q'$, and thus such a $Q$, does not exist.
However, if $Q'$ is a less-than-$k-1$-support operator, this argument does not work well.
In particular, if $Q'$ becomes a trivial conserved quantity ($H$, $I$, or 0), we cannot conclude that such a $Q$ does not exist.
In this Supplementary Material, we consider this exceptional situation.

Suppose that at least one of $C^x$s is a $k$-support operator, while the sum $\sum_{x=1}^mC^x$ is a less-than-$k-1$-support operator.
This happens, for example, when $m=2$ and $C^1=-C^2$, where we have $Q'=\sum_{x=1}^2C^x=0$.
In such a case, we consider the following sum:
\eq{
S_i^a:=\sum_{x=1}^m e^{2\pi \i ax/m} C_i^x
}
with $a=1,2,\ldots, m-1$.
It is easy to confirm that at least one of $a$ makes $S^a$ as a $k$-support operator.
By fixing $a$ to this value, we introduce a quantity
\eq{
\tilde{Q}:=\sum_{x=0}^{m-1} e^{2\pi \i ax/m}\cdot T^{(x)}Q=\sum_{i=1}^Le^{2\pi \i ai/m}S_i^a,
}
which is conserved as long as $Q$ is conserved.
In the following, we analyze $\tilde{Q}$ instead of $Q$ itself.

Although we shall treat only the case with $k=3$, the following argument is easily extended to a general $k$.
We employ the same symbol $q$ to express the coefficient in $S^a$ as $S^a=\sum_{\bsA}q_{\bsA}\bsA$.
Following a similar argument to {\it Step 1} in the main text, we find that $S^a$ should consist of doubling-product operators.
In contrast, the linear relation between these doubling-product operators is a little modified from \eref{4-support-sum}.
For example, the relation with $\x\y$ ($XZY$) and $\y\z$ ($YXZ$) provides a relation not $J_Zq_{XZY}=J_Xq_{YXZ}$ but
\eq{
J_Zq_{XZY}=e^{2\pi \i a/m}J_Xq_{YXZ}.
}
We now consider the following two sequences
\balign{
\x\y & \to \y\z \to \z\x \to \x\y \nt \\
\x\y & \to \y\x \to \x\y , \nt
}
which respectively provide
\balign{
J_Zq_{XZY}&=e^{2\pi \i a/m}J_Xq_{YXZ}=e^{4\pi \i a/m}J_Yq_{ZYX}=e^{6\pi \i a/m}J_Zq_{XZY} \\
J_Zq_{XZY}&=-e^{2\pi \i a/m}J_Zq_{YZX}=e^{4\pi \i a/m}J_Zq_{XZY}.
}
They indicate $e^{6\pi \i a/m}J_Zq_{XZY}=e^{4\pi \i a/m}J_Zq_{XZY}$.
Since $J_Z\neq 0$ and $a=1,2,\ldots, m-1$, we conclude that $q_{XZY}=0$ and thus all the coefficients of $k$-support operators are zero.

\clearpage
\end{widetext}

\end{document}